\documentclass[amssymb,pra,twocolumn,aps,showpacs]{revtex4}
\usepackage{bm}
\usepackage{graphicx}

\begin{document}

\title{Observation of bosonic coalescence of photon pairs\\}

\author{Giovanni~Di~Giuseppe,$^1$\footnote{Also at:
    Istituto Elettrotecnico Nazionale {\it G.~Ferraris}, Strada delle
    Cacce 91, I-10153 Torino, Italy.}\,\, Mete~Atat\"{u}re,$^2$\footnote{Present Address: Institute of Quantum Electronics,
    ETH H\"{o}nggerberg HPT, CH-8093 Z\"{u}rich, Switzerland.}\,\,
    Matthew~D.~Shaw,$^1$\footnote{Present Address: Stanford Linear Accelerator Center,
    2575 Sand Hill Road, Menlo Park, California 94025, USA.} Alexander~V.~Sergienko,$^{1,2}$
    Bahaa~E.~A.~Saleh,$^1$ Malvin~C.~Teich,$^{1,2}$}
\author{}
\affiliation{$^1$Department of Electrical \& Computer
Engineering,} \affiliation{$^2$Department of Physics,
         \\ Quantum Imaging
Laboratory,\footnote{URL: http://www.bu.edu/qil} Boston
University,\\ 8 Saint Mary's Street, Boston, MA 02215}

\author{Aaron~J.~Miller}

\affiliation{National Institute of Standards and
Technology\footnote{Contribution of the U.S. Government; not
subject to copyright.}, Mail Code 814, 325 Broadway, Boulder, CO
80395}

\author{Sae~Woo~Nam}

\affiliation{National Institute of Standards and
Technology\footnote{Contribution of the U.S. Government; not
subject to copyright.}, Mail Code 814, 325 Broadway, Boulder, CO
80395}

\author{John~Martinis}

\affiliation{National Institute of Standards and
Technology\footnote{Contribution of the U.S. Government; not
subject to copyright.}, Mail Code 814, 325 Broadway, Boulder, CO
80395}

\date{\today}

\begin{abstract}
Quantum theory predicts that two indistinguishable photons
incident on a beam-splitter interferometer stick together as they
exit the device (the pair emerges randomly from one port or the
other). We use a special photon-number-resolving energy detector
for a direct loophole-free observation of this
quantum-interference phenomenon. Simultaneous measurements from
two such detectors, one at each beam-splitter output port, confirm
the absence of cross-coincidences.
\end{abstract}

\pacs{42.50.Dv, 42.50.Xa, 42.50.St}

\maketitle

{\em Introduction.}---The seminal experiment carried out by Hong,
Ou, and Mandel some fifteen years ago \cite{Hong87_PRL1} is one of
the most important in the annals of quantum optics.  This
experiment demonstrated that two indistinguishable photons,
incident  on the two ports of a simple beam splitter, interfere in
such a way as to always emerge as a pair, exiting randomly from
one port {\it or} the other. The photons stuck together, so to
speak. They could only observe this  phenomenon indirectly,
however, since traditional single-photon-counting  detectors
cannot register more than one photon within the dead-time period
of the device. Hong, Ou, and Mandel circumvented this technical
limitation  by designing an experiment in which they measured the
complement of what they sought to observe. They used two
single-photon-counting detectors, placing one  at each output port
of the beam splitter, and then searched for cross-coincidences
between the two detectors. Finding none, they inferred that the
two photons do not exit from different ports of the beam splitter.
Based on this ``test of exclusion",  and energy conservation, they
deduced that both photons must exit via the same  port of the
device. There have been many variations on this original theme but
all rely on the same ``test of exclusion", whatever the source of
the single photons \cite{Rarity89,Grangier02,Santori02,Kim03}.

We have carried out a polarization version of the Hong-Ou-Mandel
experiment and report a {\it direct loophole-free observation} of
photon coincidences at a single output port of the interferometer,
thereby confirming the bosonic coalescence of pairs of photons in
a single mode \cite{Grangier02}. What makes this possible is a
unique energy detector \cite{Cabrera98} that has the capability of
registering the {\it number} of photons impinging on it. Our
experimental results serve to provide a full confirmation of the
quantum theory of photon interference in a beam-splitter
interferometer \cite{Yurke86,Fearn89,Campos89,Campos90}.

We have also made measurements at both output ports of the beam
splitter using two such detectors. This has permitted us to
demonstrate the enhancement of coincidences at a single output
port concomitantly with the diminution of cross-coincidences at
the two output ports. We thus simultaneously conduct two
experiments: a new single-port photon coincidence measurement and
the original photon cross-coincidence measurement. Our
observations confirm the inferences made by Hong, Ou, and Mandel
\cite{Hong87_PRL1} and Shih and Sergienko \cite{Shih94_PLA}.

{\it Experiment.}---The experimental arrangement, shown in Fig.~1,
is similar to that used in related quantum-interference
experiments \cite{Atature02_PRA2}, but photon-number resolving
(PNR) transition-edge sensors \cite{Cabrera98} replace the usual
photon-counting avalanche photodiodes.

\begin{figure}[ht]
   \centering
   \includegraphics[height=7cm, width=9cm]{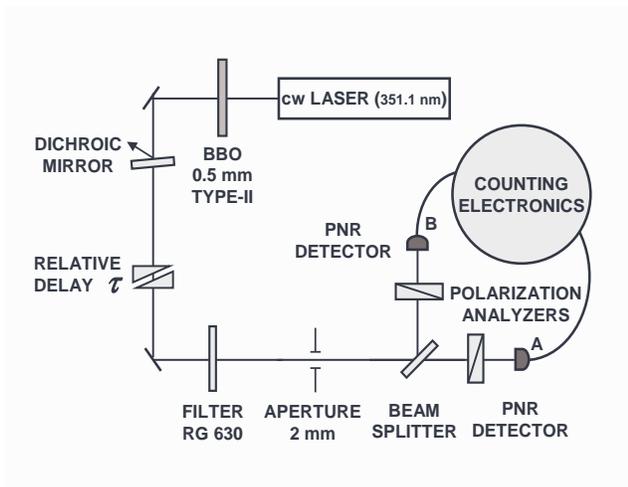}
   \caption{Schematic of the polarization-based analog of the
Hong-Ou-Mandel interference experiment using type-II collinear
degenerate spontaneous parametric down-conversion. Photon-number
resolving (PNR) detectors permit the direct observation of photon
coincidences at a single output port of the beam-splitter
interferometer as well as cross-coincidences at the two output
ports.}\label{Fig1}
\end{figure}

To generate orthogonally polarized photon pairs, a single-line
351.1-nm continuous-wave (cw) argon-ion laser operated at 100 mW
was used to pump a 0.5-mm-thick $\beta$-BaB$_2$O$_4$ (BBO)
nonlinear-optical crystal, aligned for type-II collinear
degenerate spontaneous parametric down-conversion (SPDC). The low
power of the pump ensures that the photon pairs are well separated
in time. A 2-mm aperture placed 70 cm beyond the crystal was used
to select only those photon pairs that propagate collinearly with
the pump. The pump is disposed of by use of a dichroic mirror and
an RG-630 colored glass filter.

At the heart of the interferometer is a non-polarizing beam
splitter that distributes the photon pairs into two spatial modes,
denoted A and B in Fig.~1. In each arm, a Glan-Thompson
polarization analyzer oriented at 45$^{\circ}$ renders the two
photons indistinguishable in polarization. Narrowband interference
filters were not used in the experiment.

A lens in each arm (not shown) couples the light into two 9-$\mu$m
mode-field-diameter optical fibers, each of which is connected to
a separate PNR detector operated in a cryostat. Detection events
from the two PNR detectors were recorded and analyzed using a
computer-controlled system that registers time-stamped single- and
double-photon events. Software is used to extract the coincidences
at a single detector and the cross-coincidences from the two
detectors.

The detector elements of the transition-edge sensor are
photolithographically patterned 40-nm-thick tungsten thin films
deposited on a silicon substrate \cite{Cabrera98}. The substrate
is cooled to approximately 60 mK, about half the
superconducting-to-normal transition temperature of 100 mK (the
transition width is about 1 mK). A bias voltage across the thin
film maintains the temperature in the transition region via Joule
heating. An incident photon absorbed by the tungsten film is
converted to a photoelectron, which raises the electron
temperature of the film, thereby increasing its resistance. The
time integral of the associated decrease in current, multiplied by
the bias voltage, provides the total photoelectric energy absorbed
by the thin film within its 15-$\mu$sec thermal relaxation time.
In conducting an optical experiment using light of a specified
wavelength, the number of photons incident within the thermal
relaxation time is determined by establishing the total energy
transferred to the detector within this time. Of course, energy
detectors of this kind cannot distinguish between the absorption
of two photons, each of energy $E$, and the absorption of a single
photon of energy $2E$; great care was therefore used to prevent UV
pump photons from leaking through to the detectors.

Over the range of wavelengths of interest in our experiments, the
quantum efficiency $\eta$ of these PNR detectors is approximately
20\% \cite{Miller03}, as determined via an absolute measurement
technique \cite{Klyshko80_KE,Rarity87,Czitrovszky00}. The finite
quantum efficiency obviates the possibility of determining the
incident photon number with certainty. Nevertheless, by virtue of
the invariance of the Poisson distribution to random deletion
\cite{Teich82}, it is possible to infer the photon-number
resolving capability of these detectors by measuring the number
distribution that results from excitation with weak laser light
\cite{Miller03}. The typical full-width half-maximum energy
resolution of these detectors is currently about 0.25 eV at 1.77
eV, corresponding to 100 nm at a central wavelength of 700 nm. The
signal is read out of the detector using a system that
incorporates an array of dc superconducting quantum interference
devices (SQUIDs), which operate as current-sensitive amplifiers.
To limit spurious ``pileup" counts, experiments are carried out
using a reduced singles counting rate $\sim$1000 counts/sec, an
order of magnitude lower than the achievable counting rate.

A Babinet-type compensator consisting of two parallel quartz
prisms and a fixed quartz plate, $z$-cut to eliminate transverse
birefringence, is used to modify the relative time delay $\tau$
between the two photons. The condition of indistinguishability is
imposed by overlapping the two single-photon wave packets in space
and time at the beam splitter; this is not required, in principle,
as long as distinguishing information is subsequently erased
\cite{Pittman96_PRL}.

An experiment is conducted by modifying the degree of
distinguishability (achieved by varying the delay time $\tau$ over
a range of hundreds of fsec) and tracing out three curves: the
coincidence probability at detector A, denoted $P(2,0)$; the
coincidence probability at detector B, denoted $P(0,2)$; and the
cross-coincidence probability, denoted $P(1,1)$.

The results of a typical experiment are illustrated in Fig.~2.
Never-before seen are the data for the peaks in the coincidence
probabilities, $P(2,0)$ and $P(0,2)$, at detectors A and B,
respectively. These peaks are a manifestation of excess photon
pairs at a single output port of the beam splitter. The
coincidence probability is enhanced because the photons are
identical bosons when the wave packets overlap. Were the same
experiment to be conducted with fermions, the coincidence
probabilities would be suppressed rather than enhanced
\cite{Loudon98}. The data points for the complementary
cross-coincidence probability $P(1,1)$ (diamonds) exhibit the
familiar dip associated with the quantum interference of
indistinguishable photons \cite{Atature02_PRA2}.

\begin{figure}[ht]
   \centering
   \includegraphics[height=7cm, width=9cm]{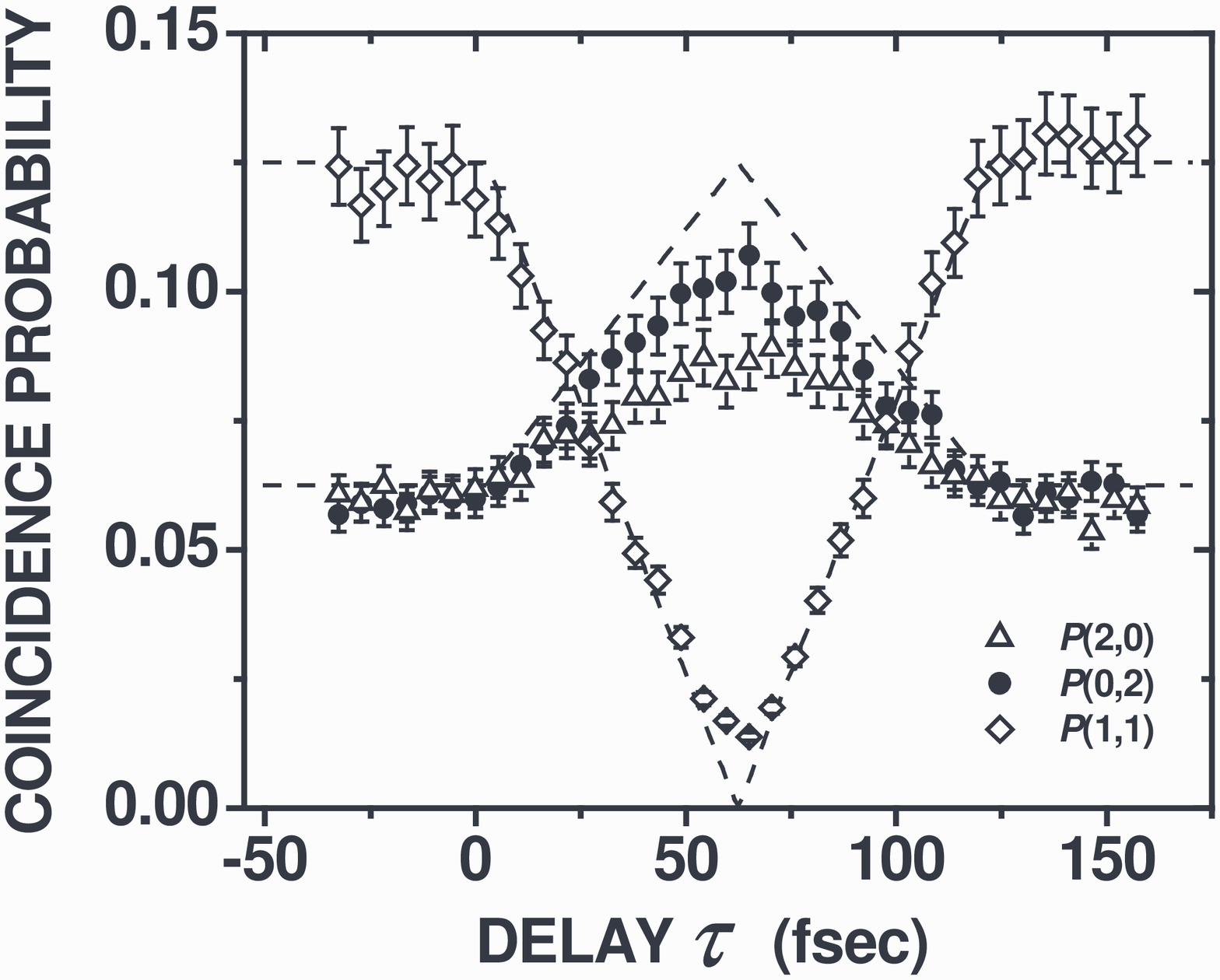}
   \caption{Experimental coincidence probabilities $P(2,0)$ and
$P(0,2)$, and cross-coincidence probability $P(1,1)$, as a
function of the relative delay time between the photons $\tau$
imparted by the compensator (symbols). The theoretical curves
(dashed) are computed from the ideal single-mode theory provided
in Eqs.~(\ref{doublesrate}) and~(\ref{coincidencerate}), assuming
a crystal of length 0.5 mm. The data follow the trends of the
theory well, but with reduced visibility resulting from imperfect
alignment and asymmetric polarization losses in the
system.}\label{Fig2}
\end{figure}

Delay times $\tau$ that are substantially larger, or smaller, than
those at the dip/peak correspond to distinguishable photons. In
that domain the coincidence probabilities are characterized by
classical particle-like statistics, namely the binomial counting
distribution \cite{Campos89}. The cross-coincidence probability is
then expected to be twice the coincidence probability, which is,
in fact, exactly what is observed on the shoulders of the
interference pattern (see Fig.~2).

{\em Theory.}---For collinear SPDC confined to a single spatial
mode, the quantum state at the output of the nonlinear crystal
driven by a monochromatic plane-wave pump at frequency
$\omega_{p}$, is \cite{Atature02_PRA2}
\begin{equation}\label{general-state}
    |\psi\rangle =
            \int d\omega\,
            {\tilde \Phi}(\omega)
            {\hat a}^\dag_o\left(\frac{\omega_p}{2}+\omega\right){\hat a}^\dag_e\left(\frac{\omega_p}{2}-\omega\right) |0\rangle
\end{equation}
\noindent where the operators ${\hat a}^\dag_o(\omega)$ and ${\hat
a}^\dag_e(\omega)$ create photons of ordinary and extraordinary
polarization, respectively, in frequency mode $\omega$, and the
limits on the integral stretch from $-\infty$ to $\infty$. The
form assumed by the state function ${\tilde \Phi}(\omega)$, which
is normalized according to $\int d\omega|{\tilde
\Phi}(\omega)|^{2}=1$ depends on the physical structure of the
down-conversion source. The state function for a single bulk
crystal of length $L$ is, for example, given by
\cite{Atature02_PRA2}
\begin{equation}\label{sinc}
    {\tilde \Phi}(\omega) \propto
        L\,{\rm sinc}\left[ \frac{L}{2}\, \Delta (\omega) \right]
        e^{i L \Delta (\omega) / 2},
\end{equation}
\noindent where the wave-vector mismatch function $\Delta
(\omega)=k_{p}(\omega_p)-k_{o}(\omega_p/2+\omega)-k_{e}(\omega_p/2-\omega)$
depends on the dispersive properties of the birefringent medium
($k_{\{p,o,e\}}$ represent the wave numbers of the pump, ordinary
wave, and extraordinary wave in the crystal, respectively).

Assuming an ideal optical system and perfect components, the
coincidence probabilities, as a function of $\tau$, are then given
by
\begin{eqnarray}\label{doublesrate}
 {P}(2,0) &=& {P}(0,2)
                    \nonumber \\
           &=& \frac{1}{32} \int dt\, \left| \Phi (t - \tau) +
                        \Phi (-t- \tau)  \right|^2,
\end{eqnarray}
\noindent where $\Phi(t)$ is the inverse Fourier transform of
${\tilde \Phi}(\omega)$ and the integration can be extended from
$-\infty$ to $\infty$ since the integrand is narrow in comparison
with the detection window. Similarly, the cross-coincidence
probability becomes \cite{Rubin94}
\begin{equation}\label{coincidencerate}
    {\rm P}(1,1) =
        \frac{1}{16} \int dt \left| {\Phi} (t - \tau) -
            {\Phi} (-t - \tau) \right|^2.
\end{equation}
The fractional pre-factors in Eqs.~(\ref{doublesrate})
and~(\ref{coincidencerate}), 1/32 and 1/16 respectively, can be
traced to the fact that the polarization analyzers in each arm of
the interferometer transmit only half the photons.

For perfect photon-wavepacket overlap, with the value of $\tau$
chosen such that $\Phi (t - \tau)=\Phi (-t - \tau)$,
Eq.~(\ref{doublesrate}) becomes $P(2,0) = P(0,2) = 2^{2}/32 = 1/8$
while Eq.~(\ref{coincidencerate}) becomes $P(1,1) = 0$, and the
result reduces to the ideal boson counting distribution.

In the opposite limit, when the relative delay time $\tau$ is
sufficiently large, the integrals in Eqs.~(\ref{doublesrate})
and~(\ref{coincidencerate}) are both equal to 2, whereupon the
classical binomial counting distribution emerges: $P(2,0) = P(0,2)
= 1/16$ and $P(1,1) = 1/8$.

More generally, Eqs.~(\ref{doublesrate})
and~(\ref{coincidencerate}) can be combined to provide a
complementary formula that relates the coincidence and
cross-coincidence probabilities for arbitrary values of $\tau$:
\begin{equation}\label{singlesdoublesrelation}
    P(1,1) + P(2,0) + P(0,2)= \textstyle\frac14.
\end{equation}
In particular, when the state function in Eq.~(\ref{sinc}) is used
in Eqs.~(\ref{doublesrate}) and~(\ref{coincidencerate}), assuming
a cw pump field and a linear approximation for $\Delta (\omega)$,
the cross-coincidence probability $P(1,1)$, as a function of
$\tau$, takes the familiar form of a triangular dip
\cite{Shih94_PLA,Atature02_PRA2} while the coincidence
probabilities $P(2,0)$ and $P(0,2)$ behave as triangular peaks.

The theoretical results for this ideal single-mode theory are
shown as the dashed curves in Fig.~2. The data follow the trends
of the theory well though the visibilities of the quantum
interference patterns are reduced below their ideal values as a
result of imperfect alignment and asymmetric polarization losses
in the optical components.

{\em Conclusion.}---We have used special photon-number-resolving
energy detectors to directly demonstrate that two
indistinguishable photons incident on a polarization analog of the
Hong-Ou-Mandel interferometer stick together as they exit the
beam-splitter ports. The absence of cross-coincidences has also
been concomitantly demonstrated. Our observations provide a full
confirmation of the quantum-optical theory of interference in a
beam-splitter interferometer. It is expected that similar results
would be observed for a non-collinear configuration. As a final
note we point out that PNR detectors, such as those used here, are
expected to find use in other quantum-optics and
quantum-information-processing experiments. Unlike their APD
counterparts, they could play a role in carrying out conclusive
tests of local realism using a beam-splitter interferometer
experiment \cite{Popescu97}. They have already been found to be
useful in a number of other important applications
\cite{Romani01}.

{\em Acknowledgments.---} We are grateful to A. Abouraddy for
valuable comments. This work was supported by the National Science
Foundation; the Center for Subsurface Sensing and Imaging Systems
(CenSSIS), an NSF Engineering Research Center; the Defense
Advanced Research Projects Agency (DARPA); and the David \& Lucile
Packard Foundation.

\bibliographystyle{apsrev}

\end{document}